\documentstyle[12pt]{article}

\input{psfig}

\newcommand{\eabe} {\begin{eqnarray}}
\newcommand{\eaen} {\end{eqnarray}}
\newcommand{\eqbe} {\begin{equation}}
\newcommand{\eqen} {\end{equation}}

\newcommand{\mrm} {\mathrm}
\newcommand{\ol} {\overline}

\newcommand{\bibl}[5]
	{#1, {\it #2} {\bf #3} (#4) #5}

\newcommand{\pair}[1] {${\rm #1 \bar #1 }$}

\begin{document}

\begin{titlepage}
\begin{flushright}
  LU TP 96-29\\
  October 1996
\end{flushright}
\vspace{25mm}
\begin{center}
  \Large
  {\bf  A Program for Baryon Generation and Its Applications to Baryon Fragmentation in DIS} \\
  \vspace{12mm}
  \normalsize
  Patrik Ed\'en\footnote{e-mail patrik@thep.lu.se}\\
  Department of Theoretical Physics\\
  University of Lund
\end{center}
\vspace{6cm}
{\bf Abstract:} \\
In an earlier paper, we discuss the ``popcorn'' model for baryon production in quark and gluon jets, and present an improved model (which we call Modified Popcorn Scenarium, MOPS). In this paper we give a manual to the MC program based on MOPS, and also discuss the application of the model to baryon fragmentation, i.e.\ fragmentation of strings originally contaning a diquark. Model predictions for baryon production in DIS are compared with data. 
\end{titlepage}

\section{Introduction} \label{sec-intro}
In ref \cite{paper}, a model for baryon production in the hadronization process is presented. The model is called ``Modified Popcorn Scenarium'' (MOPS), and is a revision of the ``popcorn'' mechanism~\cite{pop} for baryon production in the Lund string fragmentation picture~\cite{Lund}. In ref \cite{paper}, the discussion is focused on baryon production in $e^+e^-$ annihilation. However, the  model can also be applied to fragmentation of strings originally containing a diquark. Physical considerations about this extension of the model are discussed in section~\ref{sec-diq}, where also some results are presented.

A large part of this paper (section~\ref{sec-manual} and forward) is a manual to a FORTRAN source code file called ``MOPS.f'', which is an implementation of the the new model and works as a complement to JETSET$\geq 7.4$ \cite{JET}. The program is intended for those who already have some experience of the JETSET Monte Carlo. 

\section{Baryon Production} \label{sec-theory}
In the ``popcorn'' model~\cite{pop} a diquark is produced in a stepwise manner. In the string it is possible to produce a \pair{q} pair of ``wrong'' colour, which does not split the string in two singlet systems. If instead the original and new quark(antiquark) form an antitriplet(triplet) state, the colour field between the produced quark and antiquark will correspond to a triplet colour field with the same strength as the original field. Then no net force is acting on the produced quark and antiquark and in accordance with the uncertainty principle they can move freely for a time inversely proportional to their energy. If the string breaks within the colour fluctuation region, an effective diquark-antidiquark pair is produced. If several break-ups occur, there will be one (or several) intermediate mesons between the baryon and antibaryon.

As no net force is acting on the colour fluctuation quarks they appear as free particles. Therefore it was in ref~\cite{pop} expected that the probability to find such a \pair{q} pair with transverse mass $\mu_\perp$ at a separation $d$ in space is proportional to $\left|\Delta_F(d,\mu_\perp)\right|^2$. The suppression factor for producing effective diquarks and intermediate mesons of a total invariant mass $M_\perp$ in a string with energy density  $\kappa$ is thus estimated by 
\eqbe |\Delta_F(d ,\mu_\perp)|^2 \sim \exp(-2\mu_\perp M_\perp /\kappa). \label{e:pop} \eqen 

When implementing the model to a MC, the suppression factor Eq~(\ref{e:pop}) was originally represented by a set of parameters, governing production rates of different diquarks and the probability for one intermediate meson. The small probability for several ``popcorn'' mesons was neglected. In ref~\cite{paper} the model was reformulated in such a way that Eq~(\ref{e:pop}) could be used explicitly in the MC. The two parameters
 \eqbe \beta_q \equiv 2\left<\mu_{\perp \mrm q}\right>/\kappa,~~{\mrm or}~~\beta_{\mrm u}~{\mrm and}~\Delta\beta \equiv \beta_{\mrm s}-\beta_{\mrm u}, \label{e:betadef} \eqen
then govern both the diquark and the intermediate meson production. The result is a reduced number of tunable parameters.

We have assumed that the colour fluctuation quarks appear as free, massive particles. This implies that they cannot move along light-cones. Instead there is an area of possible starting points for the colour fluctuation, which is essentially given by the proper time of the production vertex squared. To get the total weight for a colour fluctuation, we must sum over all possible starting points. The result is a relative suppression of diquark vertices at early times, and consequently a softening of baryon momentum spectra, consistent with experimental data.

\section {Diquark Jet Fragmentation} \label{sec-diq}
\begin{figure}[tb]
  \hbox{
     \vbox{
	\begin{center}
	\mbox{
	\psfig{figure=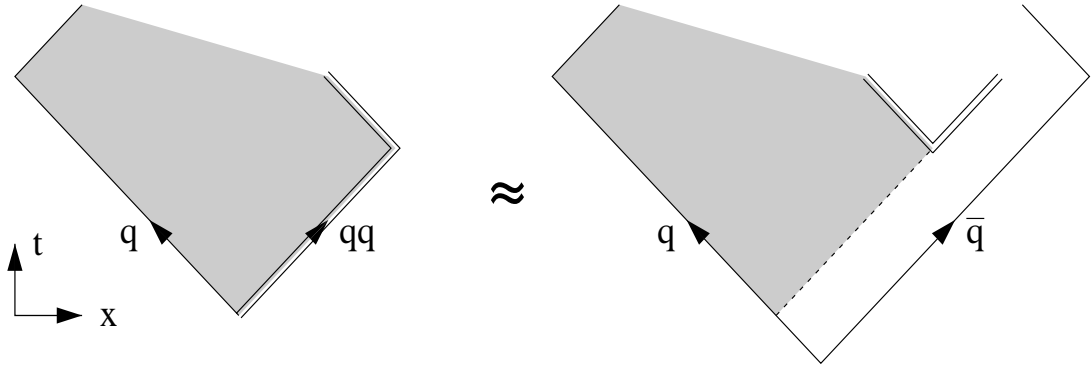,width=15cm,height=6.36cm}
	}
	\end{center}
    }
  }
  \caption{\em A string with an original diquark can be assumed to have similar properties as a string remainder after an effective diquark break-up.}
  \label{f:fig1}
\end{figure}

An algorithm for baron production in a \pair{q} string can be applied to fragmentation of a string originally containing a diquark. This string is then assumed to behave like the remainder of a \pair{q} string after an effective diquark vertex, as symbolically shown in Fig~\ref{f:fig1}.

\begin{figure}[tb]
  \hbox{
     \vbox{
	\begin{center}
	\mbox{
	\psfig{figure=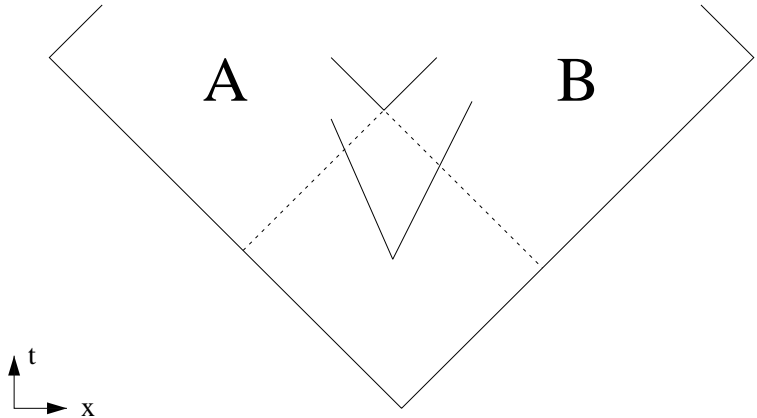,width=11.8cm,height=7cm}
	}
	\end{center}
    }
  }
  \caption{\em Two string segments A and B, causally connected by a common colour fluctuation quark pair.}
  \label{f:CommonPopcorn}
\end{figure}

This identification must be made with some care. The general statement that a remainder string behaves like an original string is based on the fact that two string systems separated by a vertex are causally disconnected. In the case of diquark vertices, this is not obviously correct. Due to the popcorn mechanism, the string segments A and B in Fig~\ref{f:CommonPopcorn} share a common set of possible histories. Choosing among those when fragmenting A will influence the probability for different final states in B.

In the specific case of the model presented in \cite{paper}, the discussion about different production points for the colour fluctuation pair, giving rise to the suppression of break-ups at small proper times (small-$\Gamma$ suppression, where $\Gamma \equiv \kappa^2\tau^2$) is not applicable to fragmentation of original diquarks. Furthermore, in the process (original diquark $\rightarrow$  baryon+$\mathrm \ol q$) the decuplet suppression should not apply at full strength. This suppression factor stems from the normalization of the overlapping q and $\mathrm \ol q$ wavefunctions in a newly produced \pair{q} pair \cite{HBN,Lund}, but in the process considered here, two out of three valence quarks already exist as an initial condition of the string.

Finally the suppression factor for popcorn mesons (which is exponential in the invariant mass of the meson system) is derived from the assumption of colour fluctuations violating energy conservation and thus being suppressed by the Heisenberg uncertainty principle. When splitting an original diquark into two more independent quarks, the same kind of energy shift does not obviously emerge. We could still expect large separations of the diquark constituents to be suppressed, since the two quarks originate from a common diquark wave function. The shape of this suppression is beyond the scope of our model, but for simplicity we assume the same kind of exponential suppression as in the "true popcorn" case. However, there is little reason to assume that the strength of the suppression must be exactly the same in the different situations. Thus the suppression of leading rank meson production in a diquark jet is governed by a new $\beta$ parameter, which is independent of the popcorn parameters $\beta_{\mathrm u}$ and $\Delta \beta$.

To summarize, when implementing the popcorn algorithm to the case of original diquarks, we have made the following changes: \begin{description}
\item[*] The $\Gamma$ suppression does not apply.
\item[*] The decuplet suppression does not apply when qq$\rightarrow$ (rank 1 B)+$\mathrm  \ol q$, but when  qq$\rightarrow$ M+...+(higher rank B)+$\mathrm  \ol q$.
\item[*] The $\beta$ parameter used in the exponential popcorn suppression is independent of $\beta_{\mathrm u}$ and $\Delta\beta$, which are fitted to $e^+e^-$ baryon production data. \end{description}

To test the model, we have made simulations of $\pi^+$p and $\mathrm K^+$p collisions at $\sqrt{s}\sim 20\mathrm GeV$, and compared the predictions with the results of ref \cite{exp}. In the simulations, we have used FRITIOF 7.02 and JETSET 7.4 (default and modified). Before presenting the results, we want to stress that the low energy-scale involved ($\sqrt{s}\sim 20\mathrm GeV \Rightarrow$ string energies mostly $< 10$ GeV) makes it dangerous to draw too firm conclusions from iterative Monte Carlo simulation programs.

\begin{figure}[htb]
  \hbox{
     \vbox{
	\begin{center}
	\mbox{
	\psfig{figure=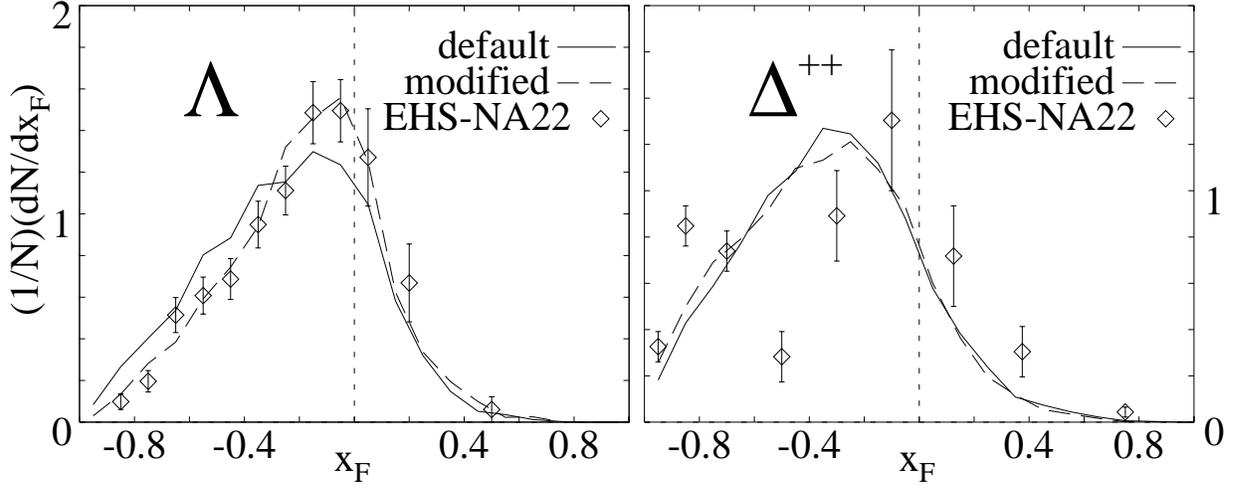,width=15.05cm,height=6.3cm}
	}
	\end{center}
    }
  }
  \caption{\em Normalized Feynman-$x$ distribution of $\Lambda$ and $\Delta^{++}$ in $\pi^+${\em p} collisions at $\sim$20GeV cms energy. The meson is moving in the positive direction. Data from \protect\cite{exp} are compared to simulations with FRITIOF7.02 combined with JETSET default (solid lines) and modified with $\beta=0.6$GeV$^{-1}$ (dashed lines).}
  \label{f:result}
\end{figure}

The results from this preliminary study are briefly (cf Fig~\ref{f:result} and Table~\ref{tab:result}):
\begin{description}
\item[*] The new model improves fit to $\Lambda$ $x_F$ spectra.
\item[*] No version reproduces the "dip" in the $\Delta^{++}$ spectra.
\item[*] The new model does not reproduce multiplicities as well as the default version.
\item[*] Changing the s/u parameter from 0.3 to 0.2, both versions improve multiplicity predictions, and now the new model is slightly better than default.
 \end{description}

The last result is of some interest, since HERA data at much higher energies also indicate support for s/u$\sim 0.2$ in DIS \cite{HERA}.

\begin{table}[htb]
\begin{center}
\begin{tabular}{||c|c|l|l|l|l||} 
  \hline
 & & \multicolumn{2}{c|}{s/u=0.3} & \multicolumn{2}{c||} {s/u=0.2} \\
baryon & EHS-NA22 & default & mod. & default & mod. \\
  \hline
$\Lambda$ & 0.092$\pm$0.008 & 0.115** & 0.152*** & 0.077* & 0.108* \\
$\Delta^{++}$ & 0.214$\pm$0.034 & 0.155* & 0.142** & 0.170* & 0.148* \\
$\Sigma^{*+}$ & 0.021$\pm$0.009 & 0.017 & 0.020 & 0.013 & 0.015 \\
$\ol \Lambda$ & 0.061$\pm$0.009 & 0.087** & 0.074* & 0.078* & 0.065 \\
  \hline
\end{tabular}
\end{center}
\caption{\em Multiplicities of baryons in $K^+$p collisions at $\sim 20$ GeV cms energy.  Data from \protect\cite{exp} are compared to simulations with FRITIOF7.02 combined with JETSET default and modified, for two different values of the strangeness suppression parameter. $N$ stars indicates that the simulated result is more than $N$ standard deviations away from experimental data.}
\label{tab:result}
\end{table}

\section{The MOPS Program} \label{sec-manual}
{\bf Installation}\\
All changes of the code are inside the subroutine LUSTRF. The modifications have been tested with JETSET version 7.403. However, the file MOPS.f does not contain the complete JETSET code, but only the modified LUSTRF (and some new subroutines called by it). Thus it should be possible to combine the modifications with different sub-versions of JETSET 7.4 and perhaps later versions. To use the modified program, remove the subroutine LUSTRF from a copy of the JETSET code. Create object files both of this copy of JETSET and of MOPS.f, with the same options you normally use when compiling JETSET, and link your main program together with both these object files. The modified program should then work exactly as JETSET $\geq 7.4$ with all default options available and some new options added.

Two precautions must be made:\begin{description}\item[* ]The new code uses some of the available positions in the LUDAT1 commonblock.\\{\em If anyone of these has been occupied in your version of JETSET it might be impossible to use the modifications.}\item[* ]Positions in a commonblock that are unused by JETSET default are given the value 0.\\{\em Thus the new switches and variables must be explicitly initiated by the user.} \end{description}

{\bf Uncovered physical aspects}\\
This is not an official change of the LUSTRF code. It is not prepared for every possible string configuration. Three known shortcomings are:
\begin{description}
\item[* ]Present implementation seriously underestimates baryon production in closed strings (e.g.\ $\Upsilon$~decay or closed strings from colour reconnection). 
\item[* ]Tensor meson production inside a colour fluctuation is not implemented.
\item[* ]Modification not available for ``junction strings'', coming from multiple interactions. (This is however a feature not fully implemented in JETSET at present.)
\end{description}
There are no immediate plans of extending the implementation to cover the points above, unless feed-back shows a need to do so.

{\bf Diquark probability shift}\\
In the way the new model is implemented, some of the produced baryons are rejected due to $\Gamma$ suppression. To get a proper amount of final baryons, the program therefore needs a larger $P({\rm qq})/P({\rm q})$ value as input. This could of course be achieved simply by changing PARJ(1). However, the default flavour selection subroutine, LUKFDI, is sometimes called by LUDECY (which handles particle decays). Since the new model is not implemented in LUDECY, a drastic enhancement of PARJ(1) could lead to an overestimate of heavy hadrons decaying into baryon-antibaryon pairs.

Instead of including a modified LUDECY in MOPS.f, the shift of $P({\rm qq})/P({\rm q})$ is implemented as a new parameter, giving a multiplicative factor to PARJ(1) when using the $\Gamma$ suppression algorithm. This new parameter has however no physical interpretation, and could be excluded by implementing modifications to LUDECY.

\section{Subroutines}
Some new subroutines are called inside LUSTRF, but the user should not need to call them explicitly. Here is however a short description.

\begin{description}
\item[SUBROUTINE LUSTRF(IP) :] fragments a string as default LUSTRF, with options added to use $\Gamma$ suppression of baryons (MSTJ(12)$\geq 4$) and new flavour algorithm (MSTJ(12)=5).

\item[SUBROUTINE POKFDI(KF1,KFQPOP,KF3,KF,PT2,PI5) :] generates a new fla\-vour and combines it with an existing flavour to give a hadron, i.e.\ the same as LUKFDI.\@ POKFDI handles incoming diquarks according to the new popcorn algorithm, switched on by MSTJ(12)=5.
	\begin{description}
	\item[KFL1 :] (input) incoming flavour.
	\item[KFQPOP :] (both) flavour of present colour fluctuation quark. When a popcorn system is started $\mrm (q \rightarrow B+qq)$, KFQPOP is initiated in POKFDI.\@ If KFQPOP is a diquark flavour, the subroutine will instead form final baryon, build up by KF1 and KFQPOP.
	\item[KFL3 :] (output) produced new flavour. Is 0 if KFQPOP is a diquark.
	\item[KF :](output)  produced hadron.
	\item[PT2 :] (input) $p_\perp^2$ of hadron to be produced.
	\item[PI5 :] (output) Transverse mass squared of produced hadron.
	\end{description}

\item[SUBROUTINE DIQINI :] initiates weights for diquark flavours, baryons and popcorn mesons in the MSTJ(12)=5 popcorn algorithm. Is called if MSTJ(5)=0 and sets MSTJ(5) to 1. {\em If changing any parameter or switch governing these weights during a run, {\em MSTJ(5)} must be reset to $0$ or {\em DIQINI} called explicitly.}

\item[SUBROUTINE PONMES(IWT) :] generates number of mesons to be produced in a popcorn system. If the total system is not accepted, a completely new one is started, and PONMES is called again.
	\begin{description}
	\item[IWT :] (input) Switch used to identify original diquark fragmentation (=1) or ordinary popcorn system (=0).
	\end{description}
	
\end{description}

\section{Parameters and Switches}
The modified program uses MSTJ(5), MSTU(121,122) and PARJ(8-10,20,27). {\em Do not combine the modification with a JETSET version in which any of these variables are occupied!}

The flavour weights depend differently on the implemented parameters in different situations, and the calculations needed to extract proper weights from the parameters can be rather lengthy. For efficiency, the program starts a simulation by calculating the large but limited number of weights. The result is stored in PARF(201-517). Strictly speaking, these positions are occupied: In JETSET, the user has an option to define flavour weights in PARF(201-1960). However, since it must be considered nonsense to combine the flavour modification of this program with ``home-made'' flavour weights, this commonblock space is used, and a combination of the new flavour algorithm and the MSTJ(15)$>0$ option is explicitly forbidden.

Note that the default values (denoted by ``(D=...)'' below) are completely nonsense! This is due to the fact that the subroutine BLOCK DATA which defines default values might change significantly between different JETSET versions. In order to make it possible to link the modification to different JETSET versions, the BLOCK DATA subroutine is not included in the file MOPS.f. Thus the default values are {\em not} related to the new model. The preliminary tune presented in \cite{paper} was: MSTJ(5)=0, MSTJ(12)=5, PARJ(8)=0.6, PARJ(9)=1.2, PARJ(10)=2.09, PARJ(20)=0.6  (from the preliminary results presented in section~\ref{sec-diq} of this paper) and PARJ(27)=0.19.

\begin{description}
\item[MSTU(121) :] (Internal) counts popcorn mesons. Each time a popcorn system is started in the new algorithm, the number of mesons to be created is stored in MSTU(121) which then is decreased by 1 for each meson produced. When MSTU(121)=0 and the popcorn system is accepted in spite of suppressions, a baryon is produced and the system completed.
\item[MSTU(122) :] (Internal) switch to remember whether a popcorn system started through $q \rightarrow B+qq$ production or a baryon remnant diquark at the string end.
\item[MSTJ(5) :] (D=0) switch to calculate different weights related to the parameters of the model. DIQINI is called by LUSTRF if MSTJ(5)=0. Is set to 1 in DIQINI.\@ Reset to 0 or call DIQINI explicitly if you change any parameters governing flavour or popcorn weights during the run.
\item[MSTJ(12) :] (D=2; suggestion 5) choice of baryon production model.
	\begin{description}
	\item[= 0-3 :] see manual, default version.
	\item[= 4 :] popcorn model (as =2) with $\Gamma$ suppression.
	\item[= 5 :] $\Gamma$ suppression and new popcorn algorithm, independent of PARJ(3-7,18). Instead depending on PARJ(8-9,27) (Baryon remnant fragmentation also depends on PARJ(20)).
	\end{description}
\item[PARJ(8,9) :] (D=0.,0.; suggestion 0.6,1.2) give $\beta_u$ and $\Delta \beta$ in $\mrm GeV^{-1}$ for the popcorn suppression factor.
\item[PARJ(10) :] (D=0.; suggestion 2.1) multiplicative factor to PARJ(1)=$P({\rm qq})/P({\rm q})$, used if MSTJ(12)$\geq$4, compensating for rejected popcorn systems due to $\Gamma$ suppression. PARJ(10) merely corresponds to a shift of the PARJ(1) value. With the present implementation of particle decays in JETSET, it is however important not to change PARJ(1) far from default. This could affect some decay channels of heavy hadrons in an undesired way.
\item[PARJ(20) :] (D=0.; suggestion 0.6) gives $\beta$ in $\mrm GeV^{-1}$ for the case of fragmenting a string spanned by an original diquark.
\item[PARJ(27) :] (D=0.; suggestion 0.19) gives the suppression of baryons in the decuplet. PARJ(27) in the new model is exactly the same as PARJ(18) in the default one. The reason for using a new spot in the commonblock is again that a change of the flavour parameters used in LUKFDI affects decay channels of heavy hadrons.
\end{description}

\Large
{\bf Acknowledgments } \normalsize\\
The Modified Popcorn Scenarium, including the extension described in section~\ref{sec-diq}, was developed together with G\"osta Gustafson. In writing the MC program, I have benefited greatly from discussions with Torbj\"orn Sj\"ostrand. I want to sincerely thank them both.


\begin{thebibliography}{99}
\bibitem{paper}
  {P. Ed\'en, G. Gustafson}, {\em LU TP 96-20} (1996), {\em to be published in Z. Phys. C}
\bibitem{pop}
  \bibl{B. Andersson, G. Gustafson, T. Sj\"ostrand}{Phys. Scripta}{32}{1985}{574}
\bibitem{Lund}
  \bibl{B. Andersson, G. Gustafson, G.~Ingelman, T.~Sj\"ostrand}
	{Phys. Rep.} {97} {1983} {31}
\bibitem{JET}
  \bibl{M. Bengtsson, T. Sj\"ostrand}{Comp. Phys. Comm.}{39}{1986}{347}\\
  \bibl{T. Sj\"ostrand}{Comp. Phys. Comm.}{82}{1994}{74}
\bibitem{HBN}
  \bibl{H. Bohr, H.B. Nielsen}{NBI-HE-78-3} {} {1978} {}
\bibitem{exp}
  \bibl{EHS-NA22 collaboration, I.V. Ajinenko et al.}{Z. Phys.}{C44}{1989}{573}
\bibitem{HERA}
  \bibl{ZEUS collaboration, M. Derrick et al.}{Z. Phys.}{C68}{1995}{29}\\
  {H1 collaboration, }{\em EPS-0479, presented at the 1995 EPS-HEP meeting, Brussels, Belgium.}
\end{thebibliography}
\end{document}